\documentstyle[12pt] {article}
\begin {document}
\parindent=15pt
\begin{flushright}
{\bf US-FT/30-95}
\end{flushright}
\vskip .8 truecm
\begin{center}
{\bf F}$_2${\bf /xG(x,Q}$^2${\bf ) RATIO AT SMALL  $x$ AND THE ROLE OF THE
 CONTRIBUTION FROM INFRARED REGION}\\
\vspace{.5cm}
M.G.Ryskin and A.G.Shuvaev \\
\vspace{.3cm}
Petersburg Nuclear Physics Institute, \\
Gatchina, St.Petersburg 188350 Russia \\
\vspace{.3cm}
and \\
\vspace{.3cm}
Yu.M.Shabelski$^*$ \\
\vspace{.3cm}
Departamento de Fisica de Part\'{\i}culas, \\
Universidade de Santiago de Compostela,  \\
15706--Santiago de Compostela, Spain
\end{center}
\vspace{.3cm}
\begin{abstract}
We consider the light quark electroproduction cross section and show that
the contribution coming from the infrared ($q_T<300\, MeV$) region is not
negligible. It changes the $F_2(x,Q^2)/xG(x,Q^2)$ ratio by 20-30\% (or even
more) for small $x$ and $Q^2\sim 4\, GeV^2$.
\end{abstract}
\vspace{1.cm}

E-mail RYSKIN@THD.PNPI.SPB.RU  \\

E-mail SHABELSKI@THD.PNPI.SPB.RU \\

E-mail SHUVAEV@THD.PNPI.SPB.RU

\vspace{1.cm}

*) Permanent address: Petersburg Nuclear Physics Institute, \\
Gatchina, Sanct-Petersburg 188350, Russia.
\vskip 1. truecm

December 1995 \\
\vskip .1 truecm

{\bf US-FT/30-95}

\newpage

\section{INTRODUCTION}

In the present paper we will consider the relation between the deep inelastic
structure function $F_2(x,Q^2)$ and the gluon density $xG(x,Q^2)$ in the region
of small $x$. At very small $x$ the main contribution to $F_2$ comes from the
sea quarks and the value of DIS cross section is given by the photon-gluon
fusion subprocess Fig. 1. For large $Q^2$ in the Leading Log $Q^2$
Approximation
(LL$_Q$A) and in a physical (say, axial) gauge it is enough to deal only with
the diagram Fig. 1a. However within the LLog($\frac{1}{x}$)A, which sums up all
the terms of the type $(\alpha_s\ln \frac{1}{x})^n$, one has to take into
account the graph Fig. 1b also.

Based on these graphs Fig. 1 one can extract the value of the gluon structure
function $G(x,k^2)$ from the experiment, where the DIS structure function
$F_2(x,Q^2)$ was measured.

The question we are going to discuss here in more detail is: what are the
values of gluon kinematical variables $x_g$ and $k^2$ which correspond to the
arguments $x=x_B$ and $Q^2$ of the DIS cross section; i.e.$F_2$?

In Sect. 2 we will remind the LL$_x$A formulae \cite{RSS} for the function
$F_2$ (and $F_L$) written in the framework of the high energy
$k_t$-factorization scheme \cite{CCH}, where the gluon $k$ (see Fig. 1) is off
mass-shell.

Then by the numerical computation in Sect. 3 we will study the ratios
of the deep inelastic (Bjorken) variables ($x_B$ and $Q^2$) and the values
of the gluon's momenta ($x_g$ and $k^2$ correspondingly), which are essential
in the integrals of  Eqs. (3) and (13); and how do these ratios depend on the
form of the gluon structure function.

To answer the last question we parametrize the gluon density by the power
functions - $xG(x,k^2)\sim x^{-\lambda}\cdot (k^2)^{\gamma}$ and plot the
logarithms of the ratios $(\ln \frac{x_g}{x_G}, \ln \frac{k^2}{Q^2},
\ln \frac{q^2}{Q^2})$ as the functions of powers $\lambda$ and $\gamma$.

Even in the case of a strong scaling violation ($\gamma =\frac{1}{2}$) the
average value $<\ln \frac{Q^2}{k^2}>$ turns out to be rather large. It
corresponds to $k^2\sim 0.8 \div 2$ GeV$^2$ for $Q^2=16 \div 64$ GeV$^2$.

Thus there appears a danger that an important part of $F_2(x,Q^2)$ is
originated from the region of a small gluon's transverse momentum
(virtualities) ($k_t<1$ GeV), where the perturbative QCD approach is not
justified.

To follow the role of the infrared (small $q_t$) region we have changed also
the mass of a light quark\footnote{The heavy quark contribution to $F_2$ was
discussed in the previous paper [1].}.The value of $F_2$ increases up to the
factor of $1.5\div 2$ when instead of the constituent mass $m_q=350$ MeV one
puts the small (close to the current) mass $m_q=10$ MeV. The situation becomes
slightly better if we assume the saturation of gluon distributions at small
$x$ and $k^2$.

At very small $x$ and $k^2$ the cross section $\sigma \sim \frac{xG(x,k^2)}
{k^2}$ should tend to a constant. Thus at $k^2<k^2_0$ (for example we put here
 $k^2_0 =1$ GeV$^2$) the gluon density $xG(x,k^2)\sim k^2$; in other words at
$k^2\leq 1GeV^2$ the anomalous dimension $\gamma =1$. In this case the rapid
decrease of $G(x,k^2)$ at $k^2\rightarrow 0$ suppresses the role of the
infrared region and the result becomes more stable (under the variation of
$m_q$). This point will be discussed also in Sect. 3.

Finally we have to stress that the infrared ($q_t, k_t<0.4$ GeV) region gives
not too small contribution to the $\frac{F_2(x,Q^2)}{xG(x,Q^2)}$ ratio and one
has to remember about it, trying to extract the gluon structure function from
the $F_2(x,Q^2)$ experimental data.

\section{SMALL $x$ DIS CROSS SECTIONS}

The cross section of heavy quarks as well as light quarks (if $Q^2$ value is
large enough) electroproduction is given schematically by the graphs in Fig. 1.
The main contribution to the cross section at small $x$ is known to come from
gluons. The lower ladder blocks presents the two-dimensional gluon distribution
$\varphi(x,k^2)$  which is a function of the fraction $x$ of the longitudinal
momentum of the initial hadron and the gluon virtuality. Its distribution over
$x$ and transverse momenta $k_T$ in hadron is given in the semihard theory
\cite{GLR} by function $\varphi(x,k^2)$. It differs from the usual gluon
structure function $G(x,q^2)$:
\begin{equation}
\label{xg}
xG(x,q^2) \,=\, \frac{1}{4\sqrt{2}\,\pi^3}
\int^{q^2}_0 \varphi (x,q^2_1)\,dq_1^2 \;.
\end{equation}
Such definition of $\varphi(x,k^2)$ makes possible to treat correctly the
effects arising from gluons virtualities. The exact expression for
this function can be obtained as a solution of the evolution equation which,
contrary to the parton model case, is nonlinear due to interactions between the
partons in small $x$ region.

In what follows we use Sudakov decomposition for the produced quarks' momenta
$p_{1,2}$ through the light-like momenta of colliding particles
$p_A\simeq Q_\gamma$ and $p_B\,\, (p^2_A = p^2_B \simeq 0)$ and transverse ones
$p_{1,2T}$:
\begin{equation}
\label{1}
p_{1,2} = x_{1,2} p_B + y_{1,2} p_A + p_{1,2T} \;.
\end{equation}
The differential cross sections of a quark pair electroproduction\footnote{
$\sigma_{el,T,L}$ denotes the virtual photon-proton transverse and
longitudinal cross sections.} have the form\footnote{We
put the argument of $\alpha_S$ to be equal to gluon virtuality, which is very
close to the BLM scheme\cite{blm};(see also \cite{lrs}).}:
$$ \frac{d\sigma_{el,T,L}}{dy^*_1 dy^*_2 d^2 p_{1T}d^2
p_{2T}}\,=\,\frac{\alpha_{em}e^2_Q}{(2\pi)^4}
\frac{1}{(s)^2}\int\,d^2 k_{T} \delta (k_{T} - p_{1T} - p_{2T})\,
\delta(y_1 + y_2 - 1)$$
\begin{equation}
\label{sel}
\times\,\, \frac{\alpha_s (k^2)}{k^2}
\varphi (k^2, x)\vert M_{el,T,L}\vert^2\frac{1}{y_2} \;.
\end{equation}
Here $s = 2p_A p_B\,\,$, $k$ is the gluon transverse momentum and $y^*$ is the
quarks rapidity in the $\gamma^{\star} p$ c.m.s. frames,
\begin{equation}
\label{xy}
\begin{array}{crl}
x_1=\,\frac{m_{1T}}{\sqrt{s}}\, e^{-y^*_1}, &
x_2=\,\frac{m_{2T}}{\sqrt{s}}\, e^{-y^*_2},  &  x=x_1 + x_2 \;, \\
y_1=\, \frac{m_{1T}}{\sqrt{s}}\, e^{y^*_1}, &  y_2 =
\frac{m_{2T}}{\sqrt{s}}\, e^{y^*_2},  &  m_T=\sqrt{p^2_T+m^2} \;. \end{array}
\end{equation}
$\vert M_{el,T,L}\vert^2$ are the squares of the matrix elements\cite{RSS}
 for the cases of transverse and longitudinal electroproduction, respectively.

In LLA kinematic
\begin{equation}
\label{q1q2}
k \simeq \,xp_B + k_{T} \;,
\end{equation}
so
\begin{equation}
\label{qt}
k^2 \simeq \,- k_{T}^2 \;.
\end{equation}
(The more accurate relation is $k^2 =- \frac{k_{T}^2}{1-x}$ but we are working
in the kinematics where $x,y \sim 0$).

The matrix elements $M$ are calculated in the Born order of QCD without
standart simplifications of the parton model since in the small $x$ domain
there are no grounds for neglecting the transverse momenta of the gluon
$k_{T}$ in comparison with the quark masses and transverse momenta, especially
in the case of light quark pair production. In the axial gauge
$p^\mu_B A_\mu = 0$  the gluon propagator takes the form
$D_{\mu\nu} (k) = d_{\mu\nu} (k)/k^2,$
\begin{equation}
\label{prop}
d_{\mu\nu}(k)\, =\, \delta_{\mu\nu} -\, (k^\mu p^\nu_B \,+\, k^\nu p^\mu_B
)/(p_B k) \;.
\end{equation}
For the gluons in $t$-channel the main contribution comes from the so called
'nonsense' polarization  $g^n_{\mu\nu}$, which can be picked out by decomposing
the numerator into longitudinal and transverse parts:
\begin{equation}
\label{trans}
\delta_{\mu\nu} (k)\, =\, 2(p^\mu_B p^\nu_A +\, p^\mu_A p^\nu_B )/s\, +\,
\delta^T_{\mu\nu} \approx\, 2p^\mu_B p^\nu_A /s\,\equiv\, g^n_{\mu\nu}.
\end{equation}
The other contributions are suppressed by the powers of $s$. Since the sum of
the diagrams in Fig. 1 are gauge invariant in the LLA, the transversality
condition for the ends of gluon line enables one to replace  $p^\mu_A$  by
$-k^\mu_{1T}/x$  in expression for $g^n_{\mu\nu}$. Thus we get
\begin{equation}
\label{trans1}
d_{\mu\nu} (k)\,\, \approx\,\, -\,2\, \frac{p^\mu_B k^\nu_T}{x\,s}
\end{equation}
or
\begin{equation}
\label{trans2}
d_{\mu\nu} (k)\,\, \approx\,\, \,2\, \frac{k^\mu_T k^\nu_T}{xys}
\end{equation}
if we do such a trick for vector  $p_B$  too. Both these equations for
$d_{\mu\nu}$  can be used but for the form (\ref{trans1}) one has to modify
the gluon vertex slightly (to account for several ways of gluon emission -- see
Ref. \cite{lrs} ):
\begin{equation}
\label{geff}
\Gamma_{eff}^{\nu} =
\frac{2}{xys}\,[(xys - k_{1T}^2)\,k_{1T}^{\nu} - k_{1T}^2
k_{2T}^{\nu} + 2x\,(k_{1T}k_{2T})\,p_B^{\nu}].
\end{equation}
As a result the colliding gluons can be treated as aligned ones and their
polarization vectors are directed along the transverse momenta. Ultimately,
the nontrivial azimuthal correlations must arise between the transverse
momenta $p_{1T}$ and $p_{2T}$ of the produced quarks.

In the case of photoproduction the alignment of the gluon makes  the
distribution of the difference of the transverse momenta of the quarks
$p_T = (p_1 - p_2)_T /2$ to be proportional to $1 + cos^2 \theta$,
where $\theta$ is the angle between the vectors $p_T$ and
$k_T = (p_1 + p_2)_T /2$.

Although the situation considered here seems to be quite opposite to the
parton model there is a certain limit in which our formulae can be
transformed into parton model ones. Let us consider the case of $pp$ collisions
and assume now that the characteristic values of quarks' momenta $p_{1T}$ and
$p_{2T}$  are many times larger than the values of gluons' momenta,
$k_{1T}$ and $k_{2T}$
\begin{equation}
\label{par1}
<p_{1T}> \gg <k_{1T}> \;, \;\; <p_{2T}> \gg <k_{2T}> \;,
\end{equation}
and one can keep only lowest powers of $k_{1T}, k_{2T}$. It means that we can
put  $k_{1T} = k_{2T} = 0$ everywhere in the matrix element $M$ except the
vertices, as it was done in the parton model approach.

On the other hand, the used assumption (\ref{par1}) is not fulfilled in a
more or less realistic case. The transverse momenta of produced light quarks
should be of the order of gluon virtualities (QCD scale values).

\section{RESULTS OF CALCULATIONS AND DISCUSSION}

The function $\varphi (y,k^2)$ is unknown at small values of $k^2$ . Therefore
we rewrite the integrals over $k_2$ in Eq. (\ref{sel}) in the form
$$ \int^{\infty}_{0} \, d k^2_{T} \delta (k_{T} - p_{1T} - p_{2T})\,
\delta(y_1 + y_2 - 1)
\frac{\alpha_s (k^2)}{k^2}
\varphi (k^2, x)\vert M \vert^2 \frac{1}{y_2} = $$
\begin{equation}
\label{int}
4\sqrt{2}\,\pi^3 \delta (k_{T} - p_{1T} - p_{2T})\, \delta(y_1 + y_2 - 1)
\alpha_s (Q^2_0) xG(x,Q^2_0)
\biggl (\frac{\vert M \vert^2}{k^2}\biggr )_{k_T\rightarrow 0} \frac{1}{y_2}
\end{equation}
$$\,+\, \int^{\infty}_{Q^2_0} \, d k^2_{T}
\delta (k_{T} - p_{1T} - p_{2T})\, \delta(y_1 + y_2 - 1)
\frac{\alpha_s (k^2)}{k^2}
\varphi (k^2, x)\vert M \vert^2 \frac{1}{y_2} \;, $$
where Eq. (\ref{xg}) is used .

Using Eq. (\ref{sel}) and Eq. (\ref{int}) we perform the numerical calculation
of light quark production.

The results of computations are presented in Figs. 2-8. In Figs. 2 and 3 we
plot the values of $F_2(x,Q^2)$ coming from the sea $u+\overline{u}$ quark
contribution in the cases of MT(S-DIS) \cite{MT} and GRV HO \cite{GRV} gluon
distributions, respectively. The parameter $Q_0$ in Eq. (13) which separates
the region of 'small' and 'large' gluon momenta $k_t$ was chosen to be equal
to 1 GeV (Figs. 2a and 3a) and 2 GeV (Figs. 2b and 3b). It is seen that the
results depend rather strongly on the value of $Q_0$ as well as on the mass of
light quark and in both cases are differed from the original MT(S-DIS) (or
GRV HO) predictions for $u+\overline{u}$ sea quark contribution (which, in our
notations correspond to $Q_0^2 > Q^2$ and $m_q = 0$) and shown by solid curves
in Figs. 2 and 3. However the difference of our calculated $F_2(x,Q^2)$ values
and the original ones is not large. It means that both the procedure of the
gluon density extraction from $F_2(x,Q^2)$ experimental data and the accuracy
of our calculations are reasonably enough and there is no a large contribution
of any (say, nonperturbative) corrections.

The only explanations is the contribution of the small transverse momenta
$k_t\leq Q_0$ and $q_t\sim m_q$ to the final value of $F_2$. Even for
$Q^2 =64GeV^2$ the difference between the cases of $m_q =10$ MeV and 100 MeV is
of about 20\% (up to 30\% for GRV HO and $Q_0 =1GeV$) at $x=10^{-3}$.

To understand better what are the typical quark ($q_t$) and gluon
($k_t, x_g$) momenta, which are essential in the integrals Eq. (13), we have
chose the parametrization of gluon distribution in the form
\begin{equation}
xG(x,Q^2)=15 \biggl (\frac{x}{0.001}\biggr )^{-\lambda}
\biggl (\frac{Q^2}{20GeV^2}\biggr )^{\gamma}
\end{equation}
and calculated the mean values of logarithms $r_q =<\ln \frac{q_t^2}{Q^2}>$,
$r_k =<\ln \frac{k_t^2}{Q^2}>$ and $r_x =<\ln \frac{x_B}{x_q}>$.

The advantage of the logarithms is that they does not destroy the main power
structure of the integrand (trying to estimate directly the value of
$<\!k^2\!>$ or $<\!\frac{1}{k^2}\!>$, one faces the divergence of the integral
over $dk^2$ at large (small) $k^2_t$ for the positive $\gamma < 1$).

The ratio of the sea $u$-quark to the gluon densities are given in Fig. 4 (in
terms of $F_2(x,Q^2)$ and $xG(x,Q^2)$ structure functions) for different values
of $\gamma$ (0.2 and 0.5) and $\lambda$ (0.2 and 0.5) in Eq. (14). At very
small $x < 3\cdot 10^{-3}$ this ratio tends to a constant and does not depend
on $x$. However, even at large $\gamma = 1/2$ (the value which corresponds to
the asymptotics (extremum) of the BFKL solution) when from the formal point of
view the convergence of the integral in the infrared region should be good
enough, the dependence of the result on the light quark mass is still not
negligible ($\sim $20\% difference between the values of $F_2$ for $m_q$ = 350
MeV and 10 MeV at $x\sim 10^{-3}$ and $Q^2$ = 16 GeV$^2$).

The situation becomes slightly better (see Fig. 5) if one assumes that for
small $Q^2$ the gluon density falls down as $xG(x,Q^2) \sim Q^2$ (say,
$\gamma$ = 1 for $Q^2 <$ 1 GeV$^2$). Such a behaviour one might expect in the
case of the gluon density saturation \cite{GLR} (i.e.
$\sigma (Q^2) \sim \frac{xG(x,Q^2)}{Q^2} \rightarrow const$ at
$Q^2 \rightarrow 0$).

Unfortunately, the saturation hypothesis is also not enough to neglect
completely the small $q_t$ contribution. Increasing the quark mass from 10 MeV
to 350 MeV one diminishs the cross section more than 15\% (for example, at
$x = 10^{-3}, Q^2$ = 16 GeV$^2$, $\lambda = \gamma = 0.5$; and 20\% for
$\gamma = 0.5$ , without the saturation).

This fact is confirmed by the $\gamma$-dependence of the function $r_k$
(Fig. 7a). As it should be expected the value of $r_k$ depends slowly on
$\lambda$ but noticiably increases with $\gamma$. For a more heavy photon
($Q^2$ = 64 GeV$^2$) $r_k$ becomes smaller. It means that the typical values
of $k^2_t$ increases with $Q^2$ not as quickly as $Q^2$ (more close to
$\sqrt{Q^2}$). In the best case ($\lambda = \gamma = 0.5$ for $Q^2 >$ 1 GeV$^2$
and $\gamma = 1$ for $Q^2 <$ 1 GeV$^2$) the value of $r_k$ ($\approx -1.7$ at
$Q^2$ = 16 GeV$^2$) corresponds to $k_t^2$, essential in the integral,  of
about 3 GeV$^2$ (smaller than the usual cutoff $Q^2_0 = 4\div5$ GeV$^2$ for the
MT or MSR parametrizations \cite{MT,MSR} of the structure functions). Without
the saturation at $Q^2$ = 16 GeV$^2$ the $r_k \approx -3$; i.e. typical
$k^2 \sim$ 0.8 GeV$^2$.

In order to estimate the values of the longitudinal gluon momenta ($x_g$) we
plot in Fig. 8 the logarithms of the $x_B$ to $x_g$ ratios. The last ratios
increase with $Q^2$ as for a small $Q^2$ the transverse momenta of quarks
sometimes becomes larger than the photon virtuality ($\sqrt{Q^2}$), see Fig. 6,
and one needs a large $x_g$ of gluon to put these quarks on mass shell. In the
region of $Q^2 \approx$ 16 GeV$^2$ the value $r_x \approx -1.2$; it means that
as an average the gluon $x_g$ is 3 times larger than the Bjorken $x_B$. This
result slightly differs from the Prytz estimation \cite{Pr} where a gluon has
$x_g = 2 x_B$.

The value of $F_2(x,Q^2)$ is the sum of transverse ($F_{2T}$) and longitudinal
($F_{2L}$) contributions. As one can see from the Table, at large $Q^2$
 the last contribution is more ifrared stable.

\section{CONCLUSION}

We hope that the ratio of $F_2$ to $xG(x,Q^2)$ presented in Figs. 4, 5 as a
functions of anomalous dimention $\gamma$ and intercept $\lambda$ may be used
in future in order to extract, for example, the values of gluon density
directly from the $F_2$ structure function (in this case the values of $\gamma$
and $\lambda$ are also come from the $F_2$ data as $\gamma = d\ln F_2/d\ln Q^2$
and $\lambda = d\ln F_2/d\ln (1/x)$.

The results we have presented here are close to the standard gluon densities
extracted from the derivative $dF_2(x,Q^2)/d\ln (Q^2)$ (using the Prytz method
\cite{Pr}, or the modified one, taken into account the NLO corrections
\cite{Pr1}). Nevertheless, here we better understand the typical values of
gluon momenta we deal with and can control the role of the small $k_T$
(infrared) region.

The mean logarithms $r_q, r_k$ and $r_x$ give us an estimate of the quark
transverse momenta which are essential in the process, and the values of gluon
virtuality ($k^2$) and proton momentum fraction ($x_g$) which correspond to the
point $x_B$ and $Q^2$ for the structure function $F_2$.

A lesson one has to extract from these calculations is the fact that an
essential values of transverse momenta ($q_t$) which do contribute to the
next-to-leading order (NLO) corrections may be quite differ from the initial
$Q^2$. At not too large $Q^2$ the value of $q^2_t$ may be even larger than
$Q^2$ (in Fig. 6 $r_q > 0$ for $Q^2 \leq$ 16 GeV$^2$.

However the main danger comes from the cases where the virtuality in the
internal loops is much smaller than $Q^2$. For example when (as it was done in
\cite{EEC}) one includes in the anomalous dimension $\gamma (\omega )$ the NLO
corrections of the type $(\alpha_s/\omega )^n$ in the first non-zero (4-loops)
gluon-gluon correction $\sim (\alpha_s/\omega )^4$, he touches the region of
the internal $q^2 \ll Q^2$; down to $q^2 \leq Q^2/100$ due to the large
numerical coefficient $\delta = 14 \zeta(3)$ in the Lipatov \cite{Lip}
diffusion-like distribution
$\exp[-\ln^2(q^2/Q^2)/(4\delta y)] = \exp[-\ln^2(q^2/Q^2)/(a\cdot n)]$
where $a = \delta /ln2 \approx 24$ and $n$ is the number of loops. After this
we
use the value of the anomalous dimension $\gamma (\omega )$ (including the NLO
corrections) as a local (from the point of view of $\ln Q^2$ space)
contribution
at the given point $Q^2$.

There are no problems if $Q^2$ is very large but one can not to do this for not
too large $Q^2$ where the essential transverse momenta of intermidiate gluons
may be rather small: close to the infrared region ($Q/10\sim \Lambda_{QCD}$).

This work is supported by the Grand INTAS-93-0079.

\newpage

\begin{center}
{\bf Table 1}
\end{center}
\vspace{15pt}
Transverse and longitudinal contributions to $F_2(x,Q^2)$ for $u\overline{u}$
pair production at several points.
\begin{center}
\vskip 12pt
\begin{tabular}{|c||c||c||r|r|r|r|}\hline
 & $x$ & $m_q$, MeV & \multicolumn{2}{c|}{$Q^2$ = 4 GeV$^2$} &
\multicolumn{2}{c|}{$Q^2$ = 64 GeV$^2$}   \\ \cline{4-7}
& & & $F_{2T}$ & $F_{2L}$ & $F_{2T}$ & $F_{2L}$  \\ \hline
Without & 0.0016 & 10 & 0.126 & 0.0354 & .392 & 0.105      \\
  &    & 350 & 0.0816 & 0.0242 & .346 & 0.102     \\
saturation & 0.0064 & 10 & 0.0593 & 0.0170 & 0.186 & 0.505    \\
  &    & 350 & 0.0378 & 0.0116 & 0.163 & 0.489    \\    \hline
With & 0.0016 & 10 & 0.102 & 0.331 & 0.369 & 0.104      \\
  &   & 350 & 0.0763 & 0.0230 & 0.338 & 0.101    \\
saturation & 0.0064 & 10 & 0.0479 & 0.0159 & 0.175 & 0.0500 \\
  &   & 350 & 0.0352 & 0.0110 & 0.159 & 0.0484   \\
\hline
\end{tabular}
\end{center}
\vspace{40pt}

\newpage

\begin{center}
{\bf Figure captions}\\
\end{center}

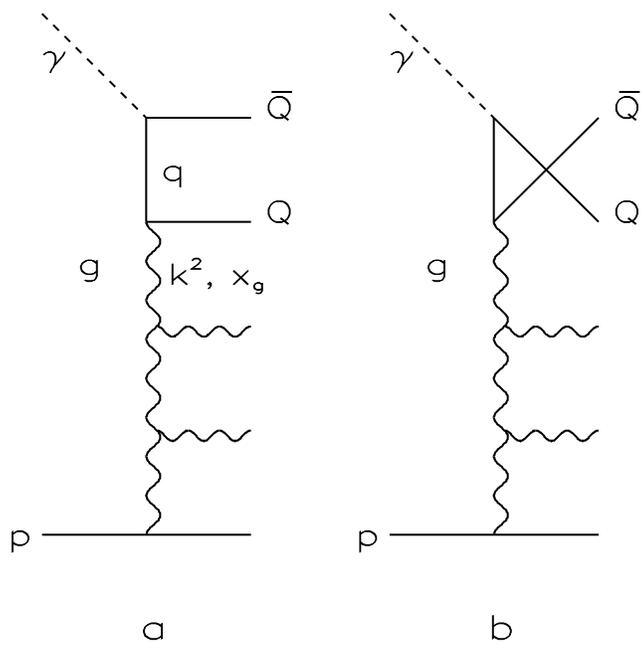
Fig. 1. Low order QCD diagrams for quark pair electroproduction via
photon-gluon fusion.

Fig. 2. $u + \overline{u}$ sea quark contributions to $F_2(x,Q^2)$ as given by
MT S-DIS \cite{MT} sea quark distributions (solid curves) and our calculations
using MT S-DIS gluon distributions for $m_q$ = 10 MeV and 100 MeV (dashed and
dash-dotted curves, respectively). The values of $Q_0$ parameter equal to 1 GeV
(a) and 2 GeV (b) were used in Eq. (13)

Fig. 3. The same as Fig. 2 but for GRV HO \cite{GRV} parton distribution.

Fig. 4. The ratios $F_2(x,Q^2)$ ($u + \overline{u}$ sea quark contribution) to
$xG(x,Q^2)$ calculated from Eqs. (13), (14) for different values $\gamma$ and
$\lambda$.

Fig. 5. The same as Fig. 4 assuming that at $Q^2 < 1$ GeV$^2$ $\gamma$ = 1
(i.e. saturation effect).

Fig. 6. $r_q =<ln\frac{q_t^2}{Q^2}>$ dependences on $\gamma$ (a) and
$\lambda$ (b).

Fig. 7. $r_k =<ln\frac{k_t^2}{Q^2}>$ dependences on $\gamma$ (a) and
$\lambda$ (b).

Fig. 8. $r_x =<ln\frac{x_B}{x_q}>$ dependences on $\gamma$ (a) and
$\lambda$ (b).

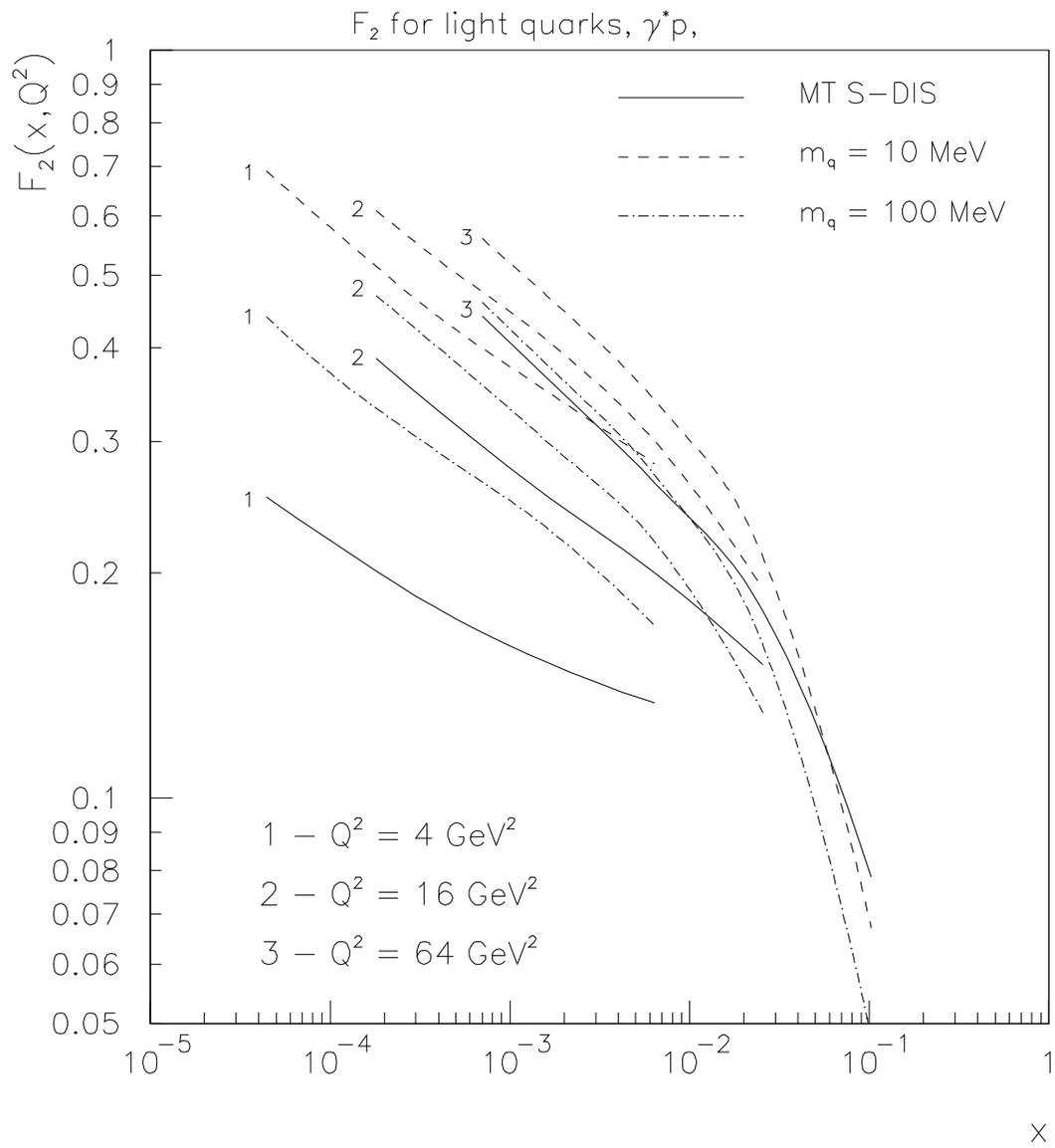
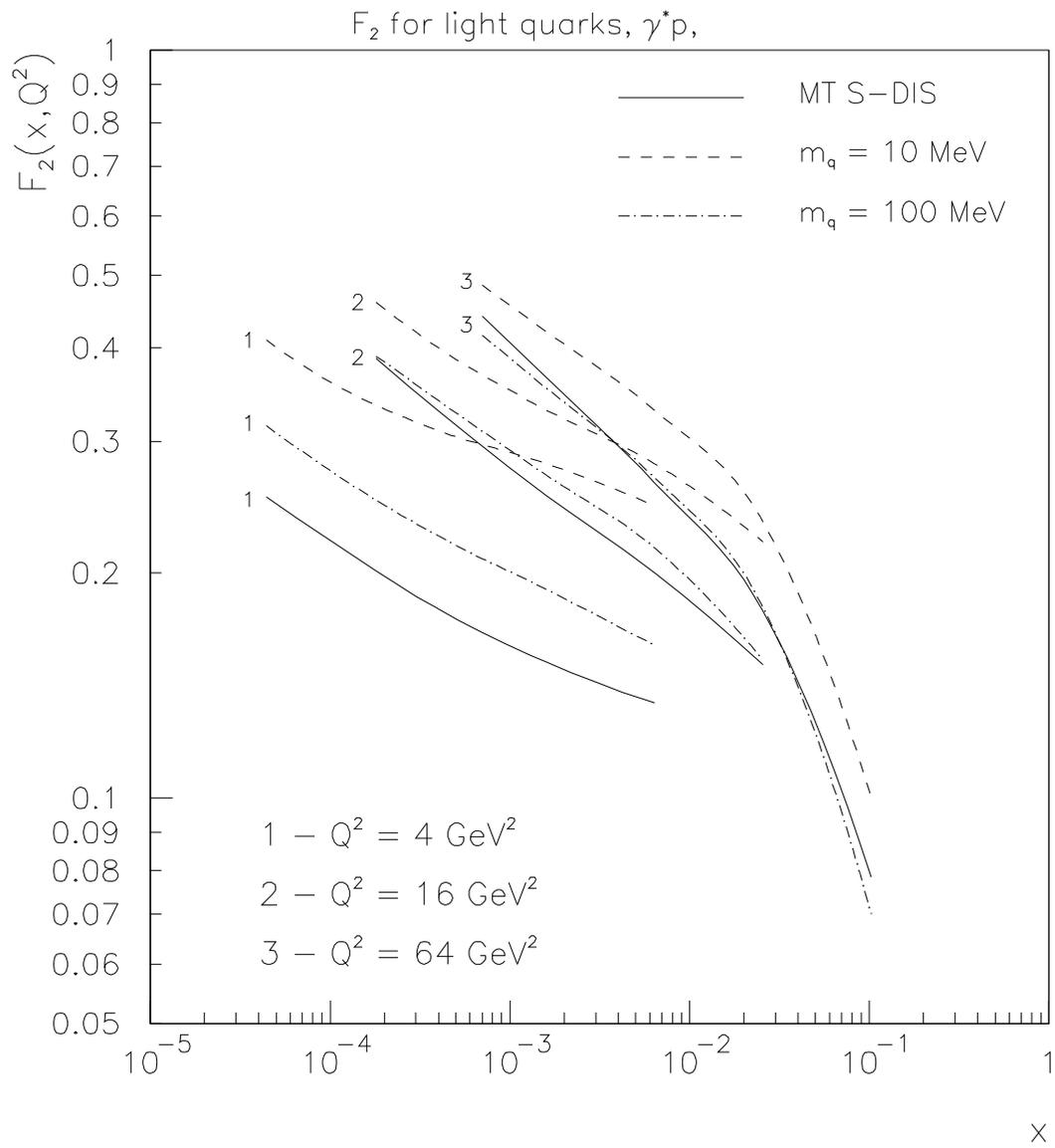
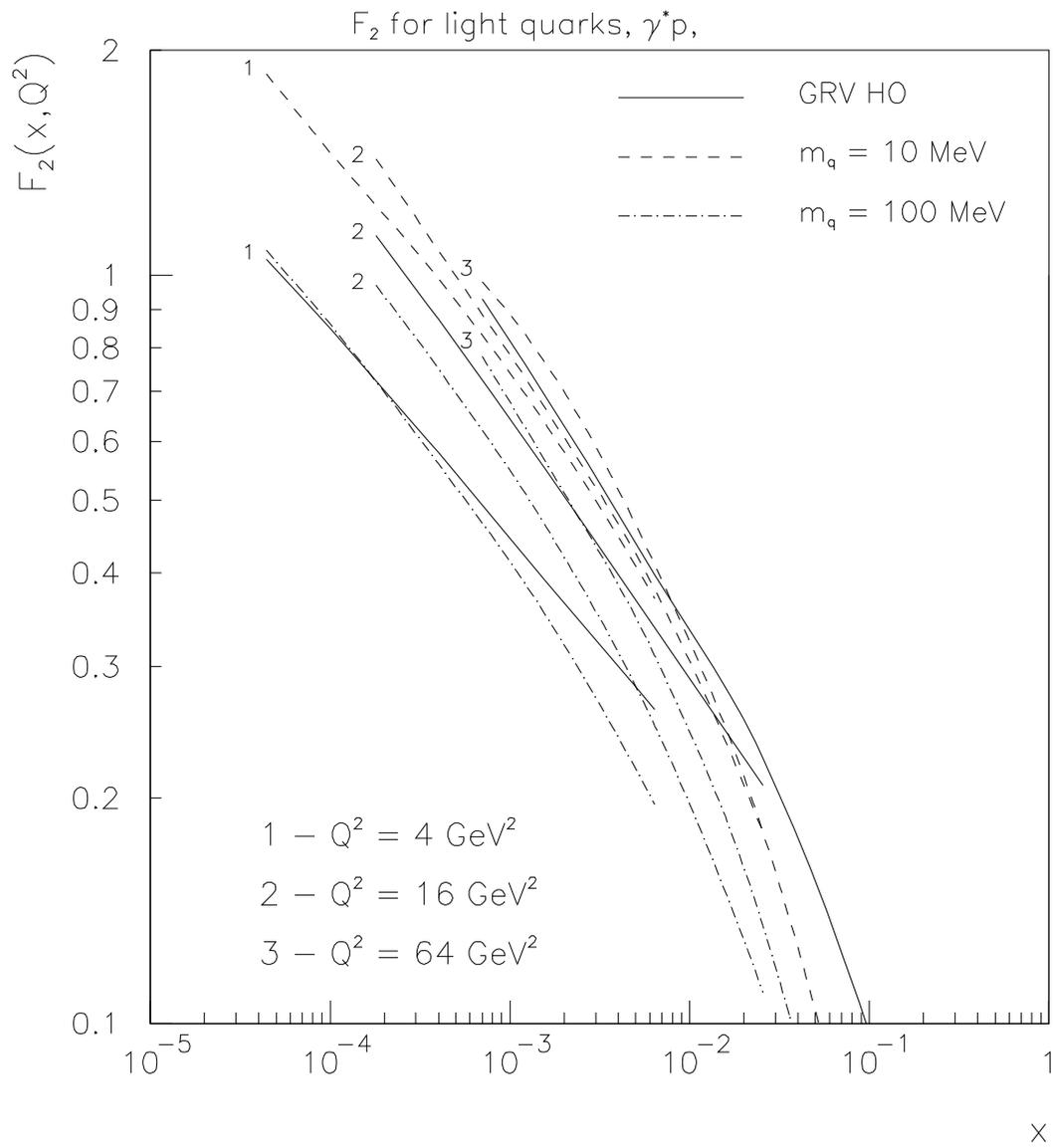
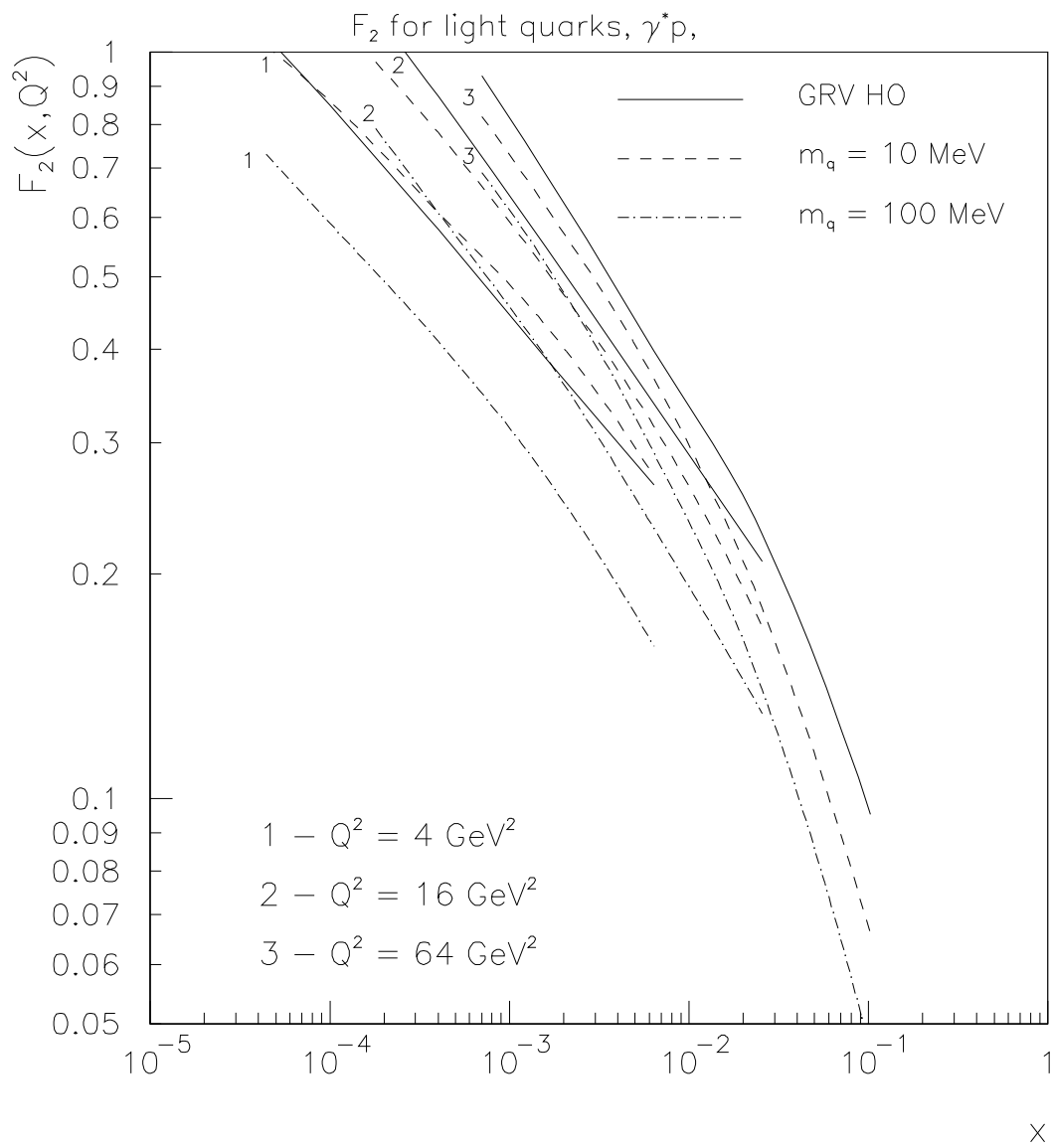
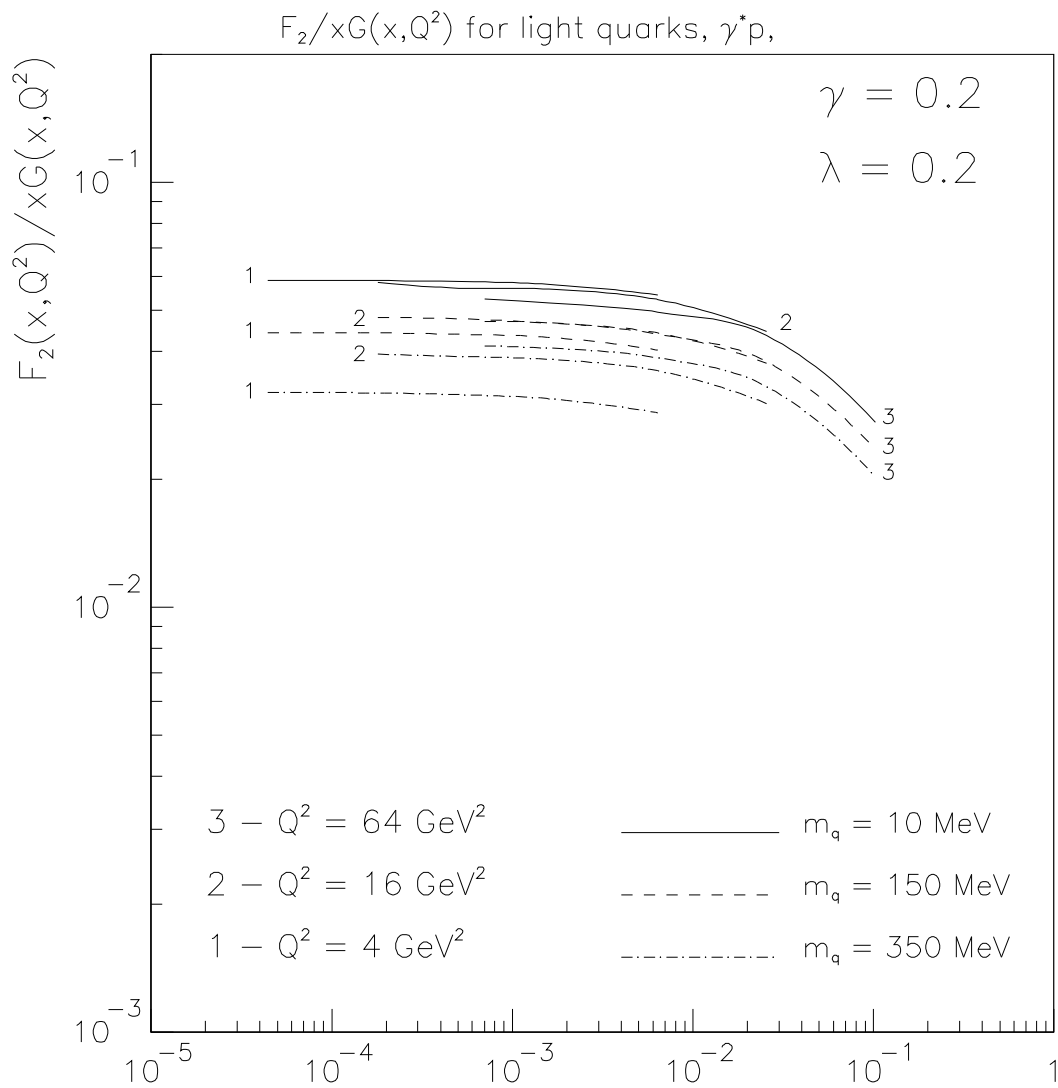
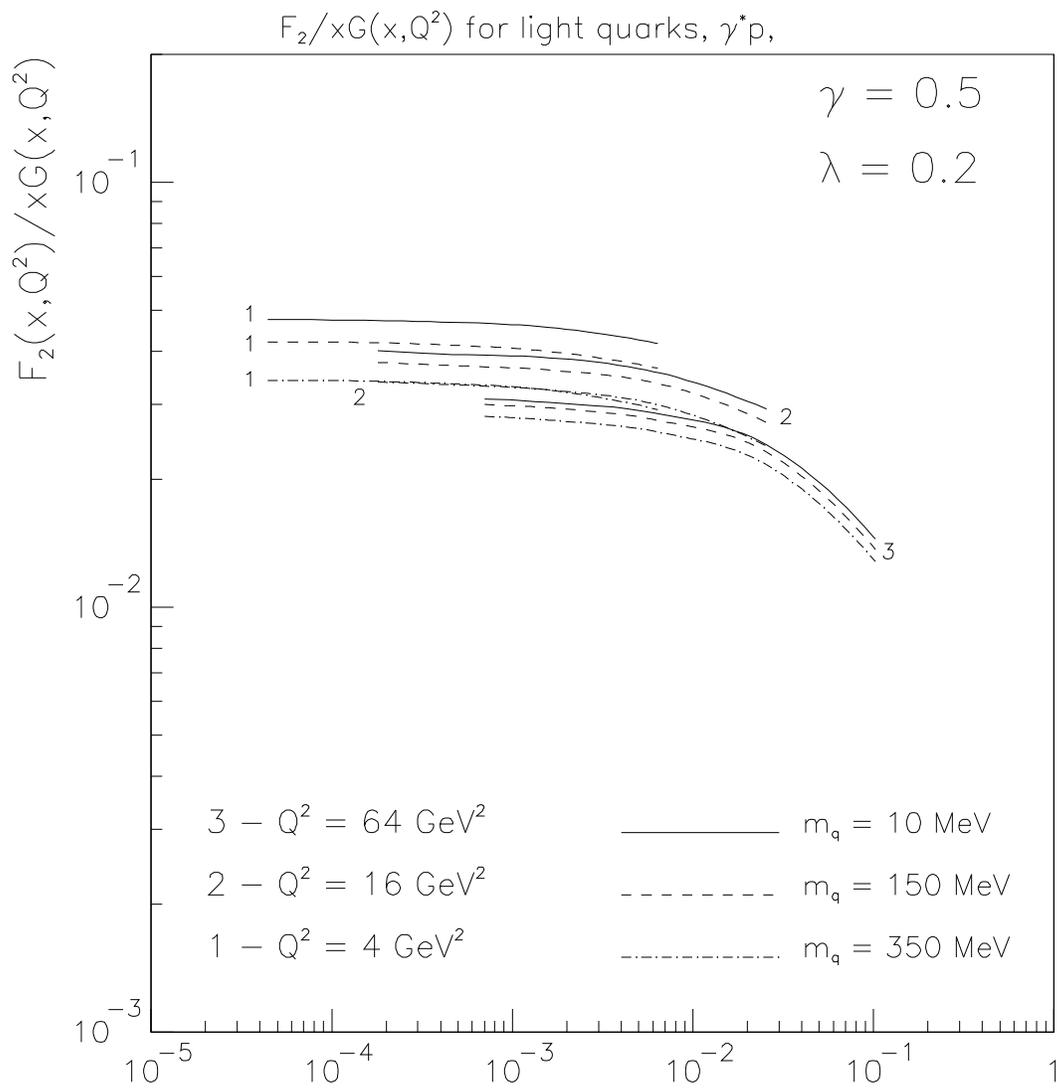
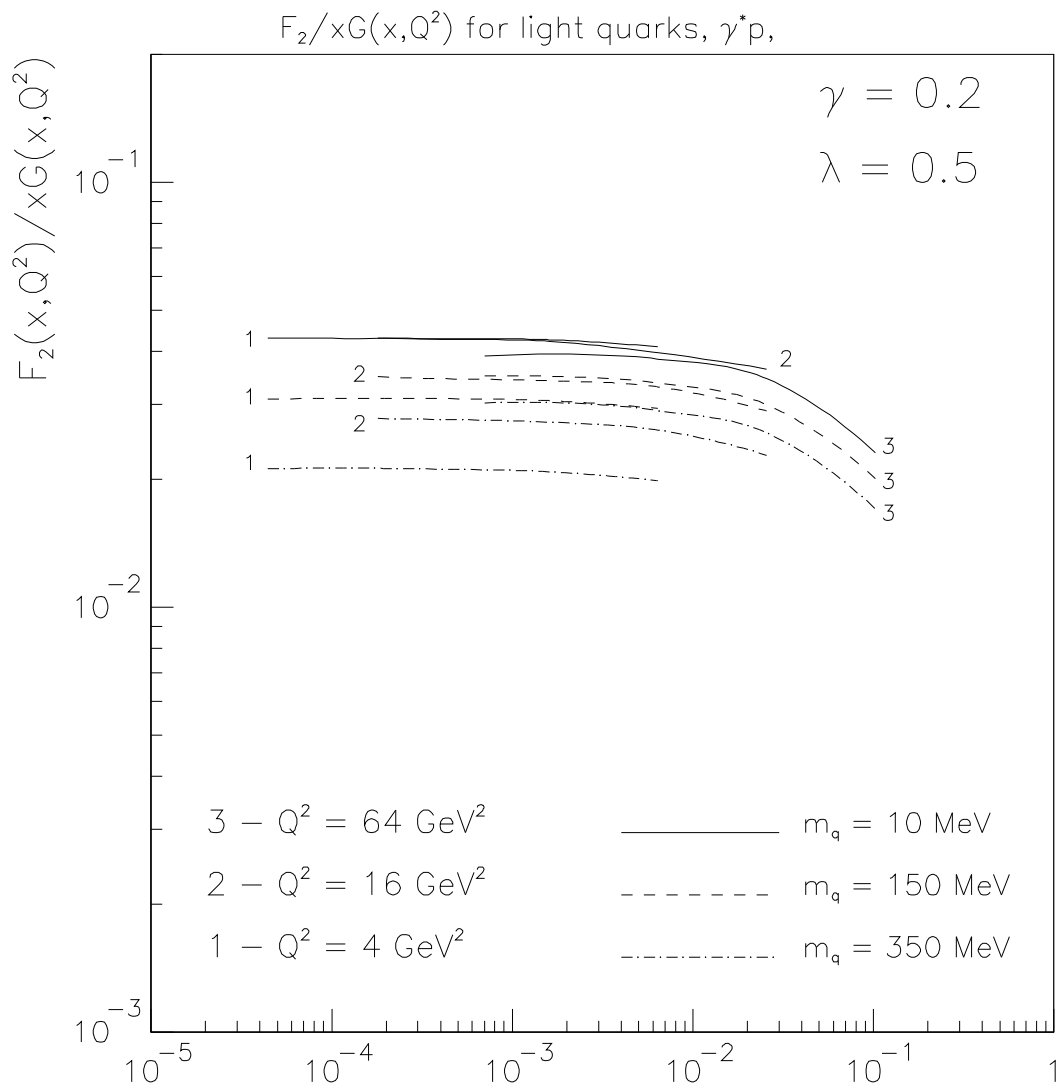
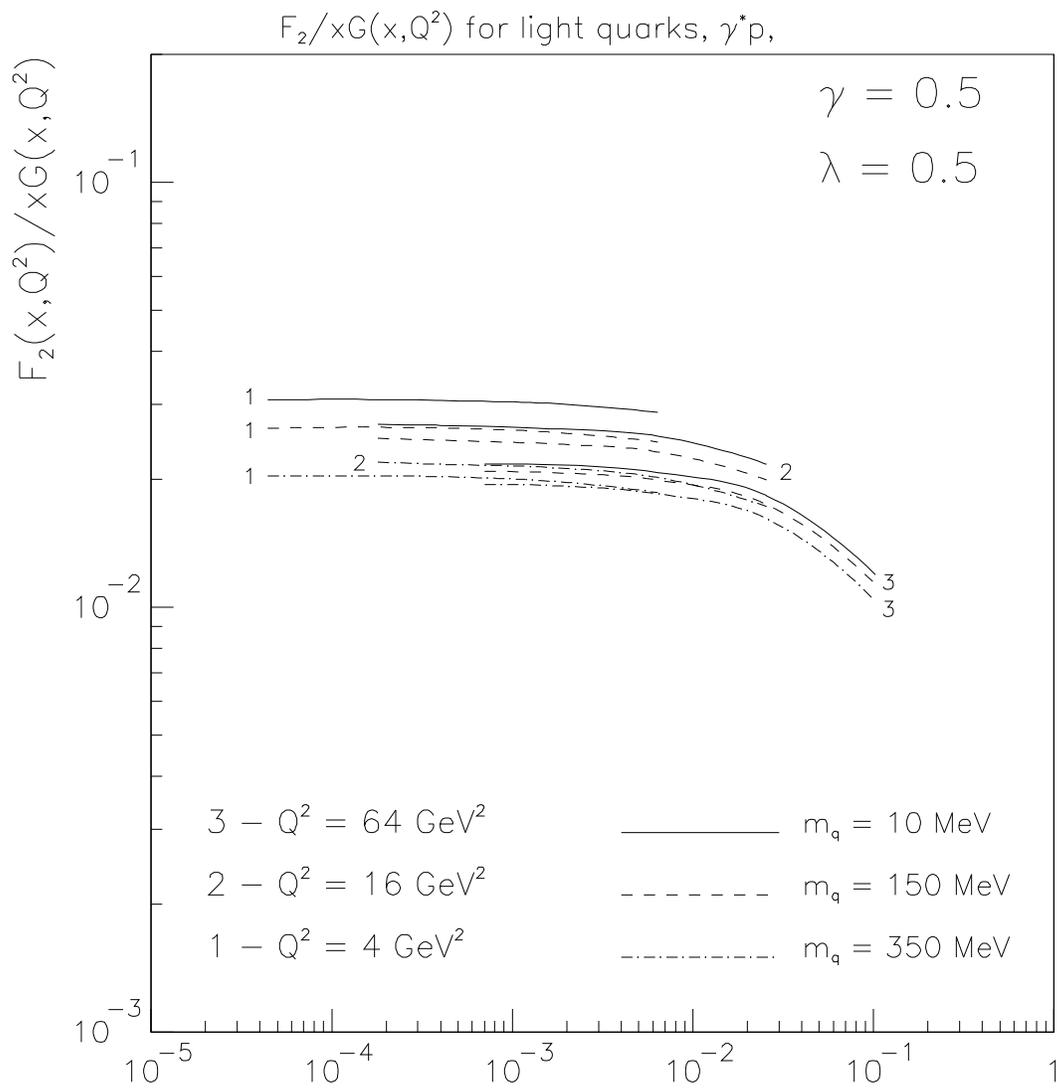
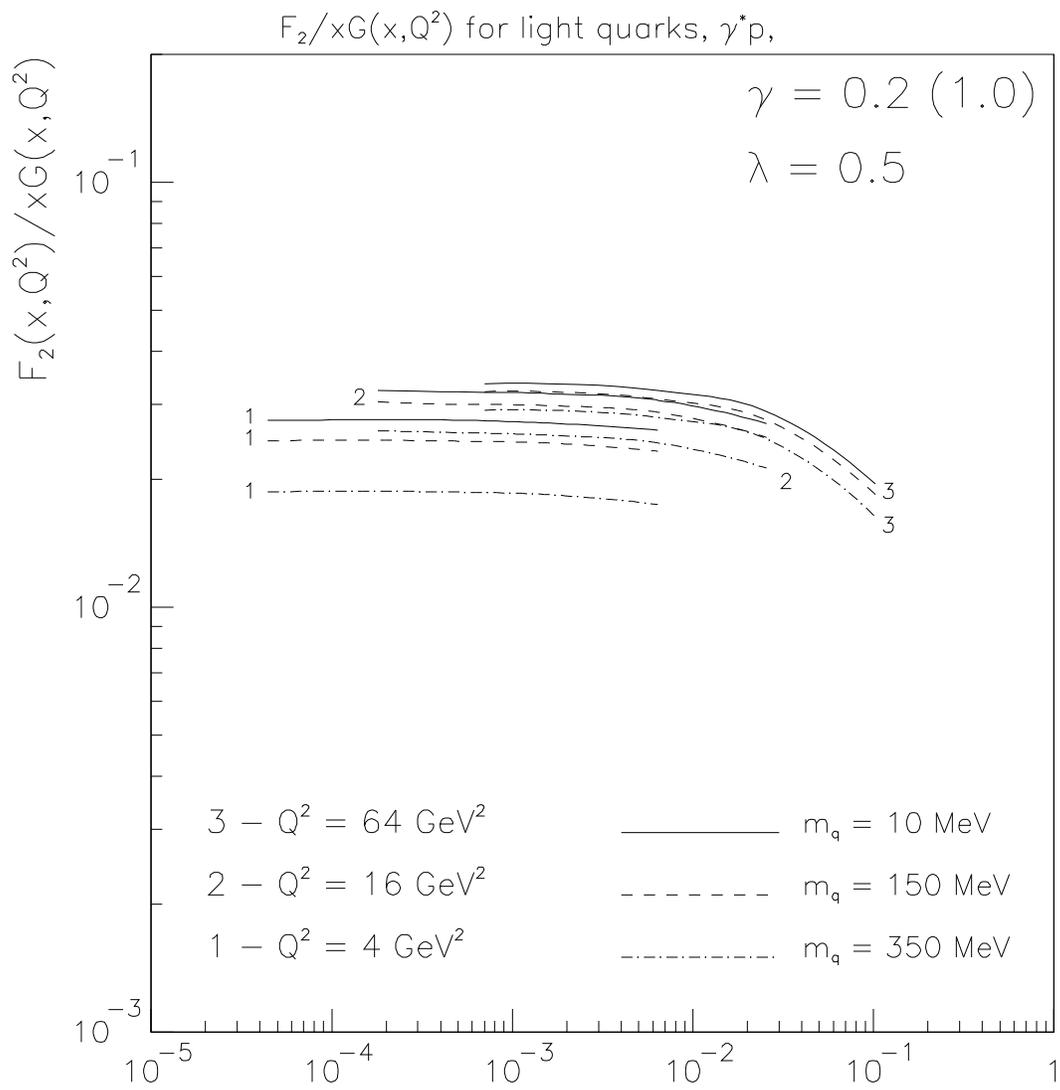
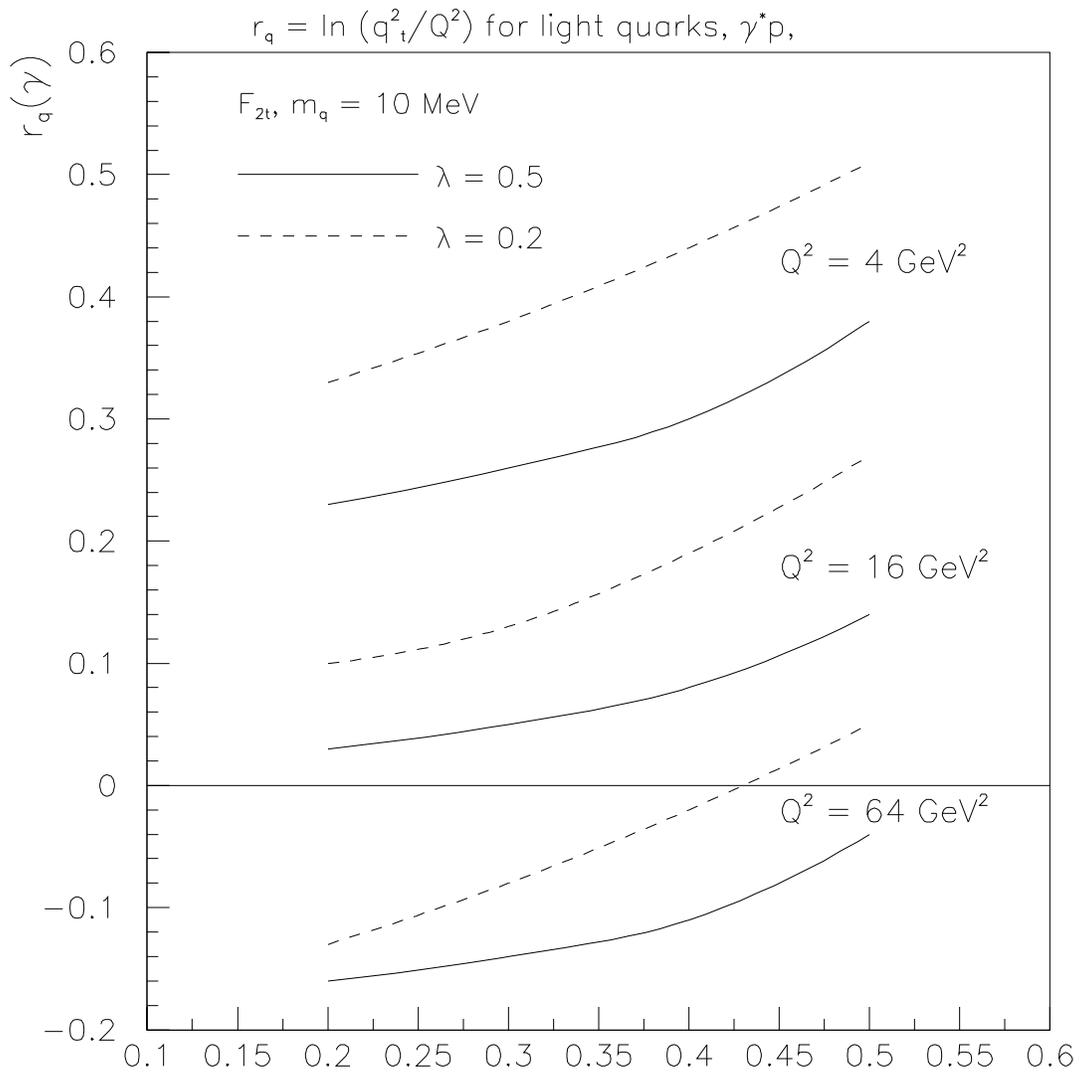
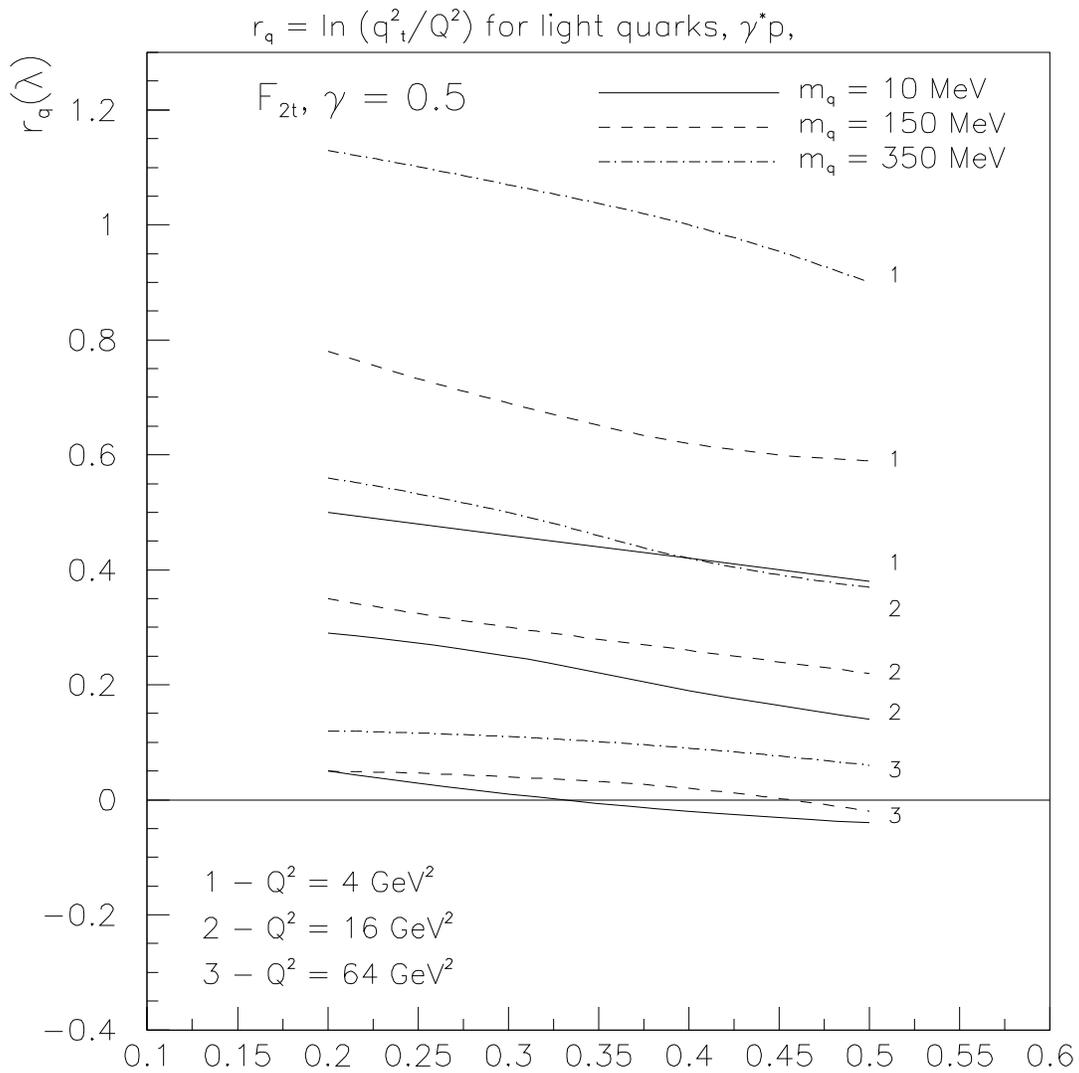
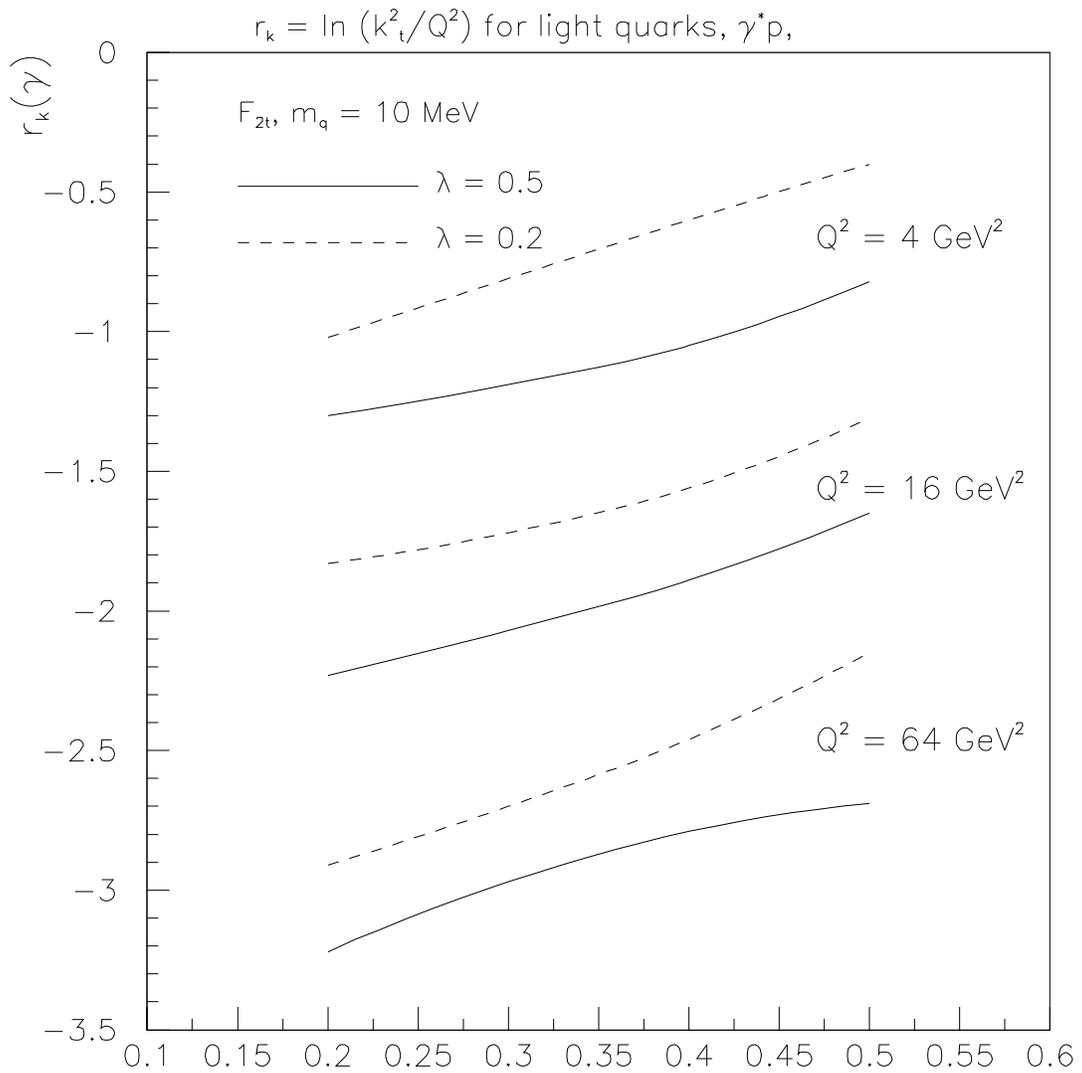
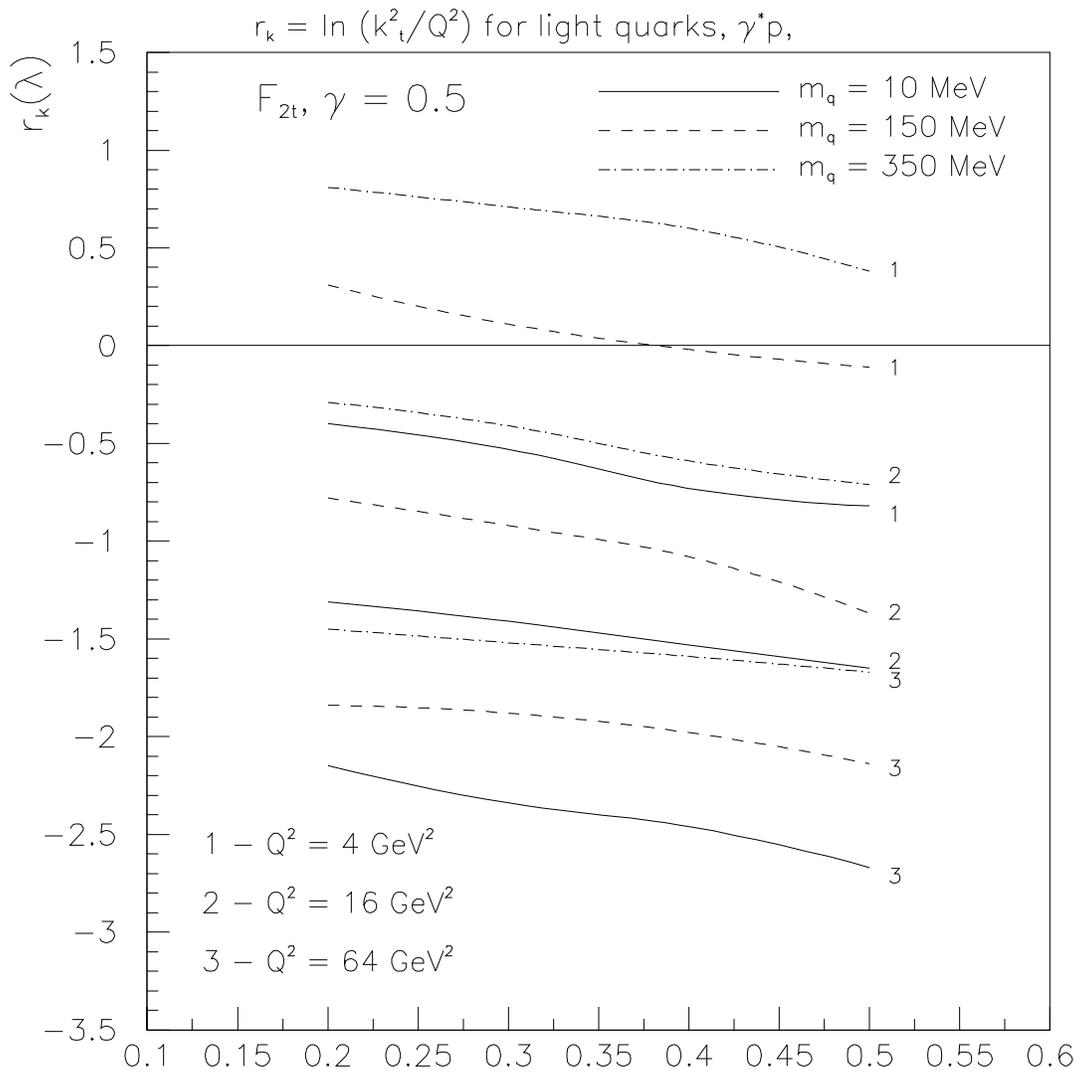
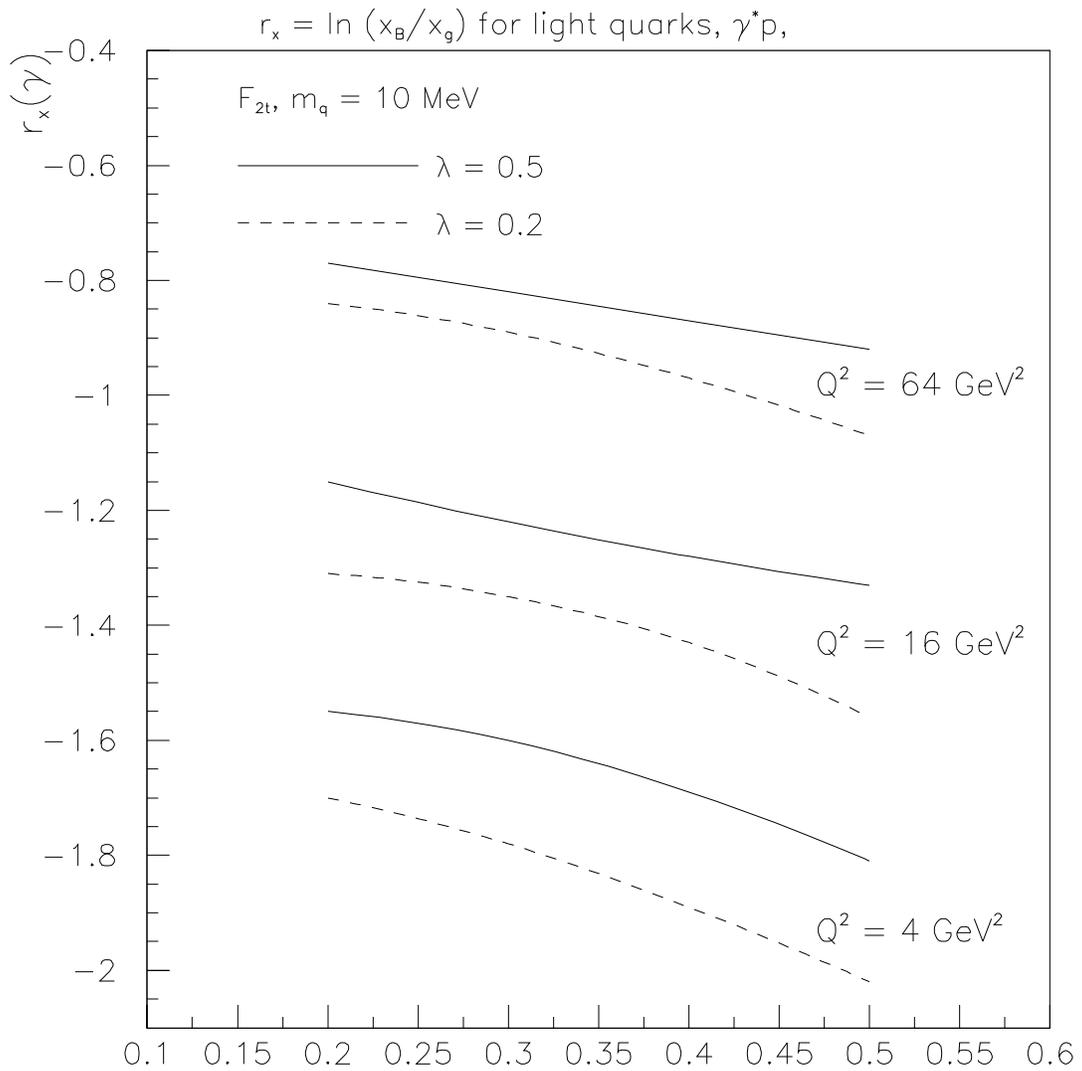
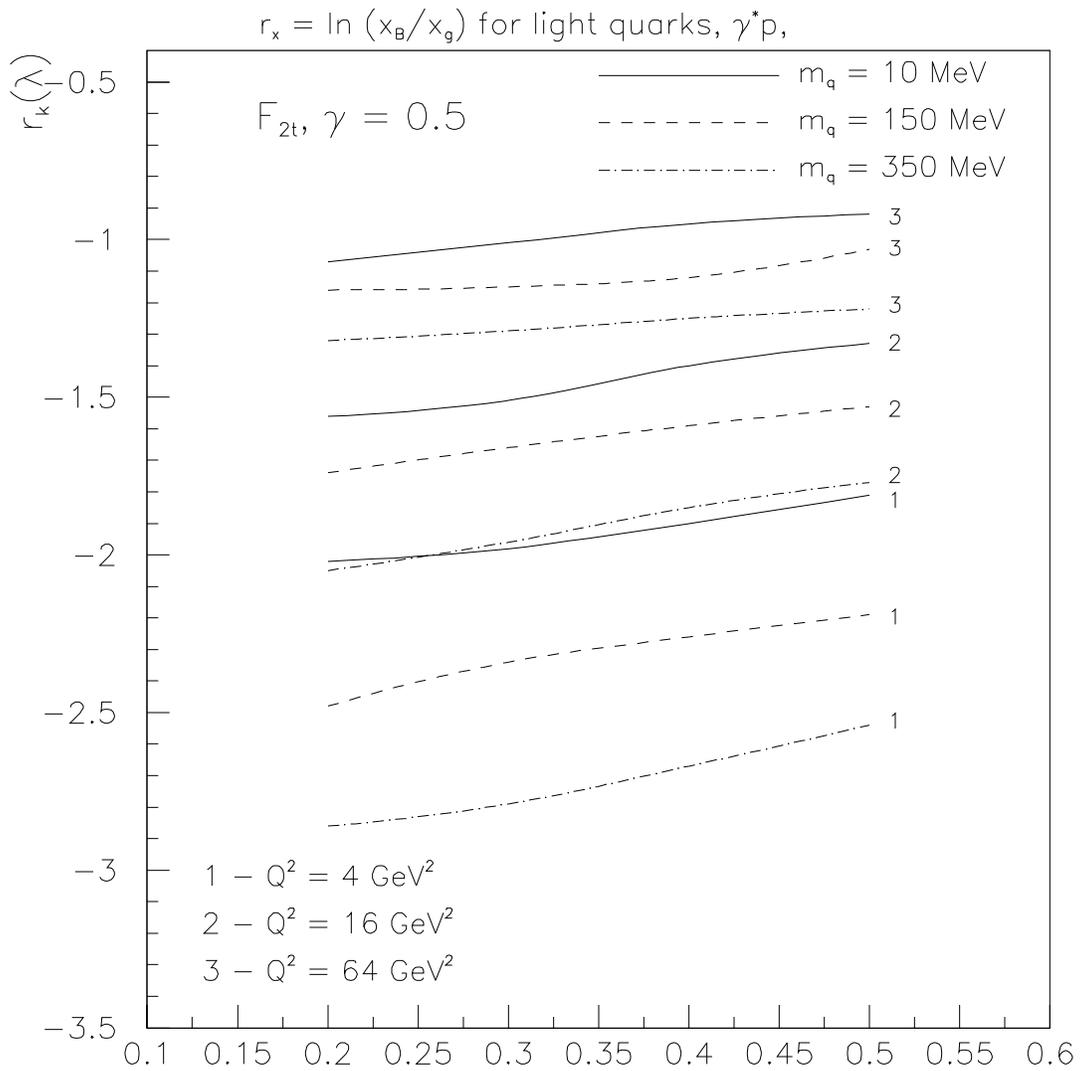
\end{document}